\documentclass[a4paper,onecolumn,11pt]{quantumarticle}
\pdfoutput=1

\usepackage{amsmath,amsthm,amssymb}
\usepackage{graphicx}
\graphicspath{{image/}}
\usepackage{bm}
\usepackage{color}
\usepackage[normalem]{ulem}
\usepackage[numbers,sort&compress]{natbib}
\usepackage{hyperref}
\usepackage{cleveref}
\usepackage{subfigure}
\usepackage{booktabs}
\usepackage{array}
\usepackage{physics}
\usepackage{tikz}
\definecolor{darkblue}{HTML}{004D6B}
\definecolor{darkred}{HTML}{8c1515}
\definecolor{darkgreen}{HTML}{006400}
\hypersetup{
    pdftitle={Sampling isometric tensor network states with monitored quantum circuits},
	colorlinks=true,
	breaklinks
}
\usetikzlibrary{calc,positioning}
\definecolor{tensorblue}{RGB}{130,183,214}
\newcommand{\rankSixTensorIcon}{%
  \begin{tikzpicture}[
    x=1cm,
    y=1cm,
    scale=0.50,
    transform shape,
    baseline=(U.center),
    tensor/.style={circle, draw=black, fill=tensorblue, thick,
      inner sep=1pt, minimum size=0.55cm},
    leg/.style={black, thick, -stealth},
    every node/.style={font=\normalsize}
  ]
    \node[tensor] (U) at (0,0) {$A$};
    \draw[leg] (-1.55,0) -- (U.180) node[pos=0, anchor=east] {$D$};
    \draw[leg] (U.0) -- (1.55,0) node[anchor=west] {$D$};
    \draw[leg] (-1.00,0.98) -- (U.135) node[pos=0, anchor=south east] {$D$};
    \draw[leg] (0,1.20) -- (U.90) node[pos=0, anchor=south] {$|0\rangle$};
    \draw[leg] (U.-90) -- (0,-1.20) node[anchor=north] {$d$};
    \draw[leg] (U.-45) -- (1.00,-0.98) node[anchor=north west] {$D$};
  \end{tikzpicture}%
}
\newcommand{\threeTensorColumnMap}{%
  \begin{tikzpicture}[
    x=1cm,
    y=1cm,
    scale=0.36,
    transform shape,
    baseline=(A2.center),
    tensor/.style={circle, draw=black, fill=tensorblue, thick,
      inner sep=1pt, minimum size=0.55cm},
    leg/.style={black, thick, -stealth},
    wrap/.style={black, thick, -stealth},
    every node/.style={font=\normalsize}
  ]
    \coordinate (A1c) at (-0.90,0.90);
    \coordinate (A2c) at (0,0);
    \coordinate (A3c) at (0.90,-0.90);

    \coordinate (A1dc) at (-0.90,-1.30);
    \coordinate (A2dc) at (0,-2.20);
    \coordinate (A3dc) at (0.90,-3.10);

    \draw[black, thick] ($(A1c)$) -- ($(A1dc)$);
    \draw[black, thick] ($(A2c)$) -- ($(A2dc)$);
    \draw[black, thick] ($(A3c)$) -- ($(A3dc)$);

    \node[tensor] (A1) at (A1c) {};
    \node[tensor] (A2) at (A2c) {};
    \node[tensor] (A3) at (A3c) {};

    \node[tensor] (Ad1) at (A1dc) {};
    \node[tensor] (Ad2) at (A2dc) {};
    \node[tensor] (Ad3) at (A3dc) {};

    \draw[leg] (-2.35,0.90) -- (A1.180);
    \draw[leg] (-1.55,0) -- (A2.180);
    \draw[leg] (-0.65,-0.90) -- (A3.180);
    \draw[leg] (A1.0) -- (1.05,0.90);
    \draw[leg] (A2.0) -- (1.75,0);
    \draw[leg] (A3.0) -- (2.45,-0.90);

    \draw[leg] (-2.35,-1.30) -- (Ad1.180);
    \draw[leg] (-1.55,-2.20) -- (Ad2.180);
    \draw[leg] (-0.65,-3.10) -- (Ad3.180);
    \draw[leg] (Ad1.0) -- (1.05,-1.30);
    \draw[leg] (Ad2.0) -- (1.75,-2.20);
    \draw[leg] (Ad3.0) -- (2.45,-3.10);

    \draw[wrap] (-2.35,1.55) .. controls (-1.40,1.55) .. (A1.135);
    \draw[wrap] (A1.-45) -- (A2.135);
    \draw[wrap] (A2.-45) -- (A3.135);
    \draw[wrap] (A3.-45) .. controls (1.40, -1.4) .. (2.50,-1.4);

    \draw[wrap] (-2.35,-.63) .. controls (-1.40,-.63) .. (Ad1.135);
    \draw[wrap] (Ad1.-45) -- (Ad2.135);
    \draw[wrap] (Ad2.-45) -- (Ad3.135);
    \draw[wrap] (Ad3.-45) .. controls (1.40, -3.60) .. (2.50,-3.60);

  \end{tikzpicture}%
}

\makeatletter
\newenvironment{inlinefigurecontent}{%
  \par\begingroup
  \def\@captype{figure}%
  \captionsetup{type=figure,hypcap=false}%
  \setcounter{subfigure}{0}%
  \centering
}{%
  \par
  \endgroup
}
\makeatother

\newcommand{\vtheta}{\boldsymbol{\theta}}

\begin{document}

\title{Sampling isometric tensor network states with monitored quantum circuits}

\author{Yuqing Rong}
\affiliation{The Hong Kong University of Science and Technology (Guangzhou), Nansha, Guangzhou, 511400, Guangdong, China}

\author{Huan-Hai Zhou}
\affiliation{The Hong Kong University of Science and Technology (Guangzhou), Nansha, Guangzhou, 511400, Guangdong, China}

\author{Guo-Yi Zhu}
\email{guoyizhu@hkust-gz.edu.cn}
\affiliation{The Hong Kong University of Science and Technology (Guangzhou), Nansha, Guangzhou, 511400, Guangdong, China}

\author{Jinguo Liu}
\email{jinguoliu@hkust-gz.edu.cn}
\affiliation{The Hong Kong University of Science and Technology (Guangzhou), Nansha, Guangzhou, 511400, Guangdong, China}

\date{}

\begin{abstract}
 
  Projected entangled pair states (PEPS) provide an efficient variational ansatz for two-dimensional quantum phases, but computing observables remains challenging because PEPS contraction is generally costly. 
  Here, we parameterize two-dimensional quantum states using variational PEPS subject to isometric constraints and map the resulting ansatz onto monitored quantum circuits, replacing tensor-network contraction with circuit sampling. 
  For infinite cylinders, the transfer matrix defines a quantum channel on the virtual boundary. We use a fixed-point treatment and a monitored-circuit unraveling of this channel to evaluate observables efficiently. 
  Using a constant number of variational parameters and a number of qubits that scales only with the cylinder width, our method yields a phase diagram for the \(J_1\)-\(J_2\) model in qualitative agreement with DMRG results. 
  Because the monitored circuits are compatible with near-term quantum hardware, this approach provides a hybrid quantum--classical framework for simulating two-dimensional quantum many-body systems.
   \end{abstract}

\maketitle

\section{Introduction}

Characterizing the ground-state properties of interacting quantum lattice models is a central problem in condensed-matter physics. 
Explicitly representing the full many-body wave function quickly becomes impractical because the Hilbert-space
dimension grows exponentially with system size, so practical methods must exploit the structure of physically relevant states. 
Tensor network states serve as a variational ansatz for many-body wave functions,
decomposing them into local tensors whose virtual bond dimensions control the
representable entanglement. 
They have become one of the main tools for describing and simulating quantum many-body systems~\cite{Orus2014,Cirac2021}. 
In one dimension, matrix product states
(MPS)~\cite{PerezGarcia2007} underlie the modern formulation of the
density matrix renormalization group
(DMRG)~\cite{White1992,Schollwock2011} and allow efficient
contraction and optimization.
In two dimensions, projected entangled pair states
(PEPS)~\cite{Verstraete2004,Cirac2021} are the natural extension
of MPS and, at fixed bond dimension, encode area-law entanglement in higher
dimensions~\cite{Verstraete2006,Eisert2010}.
The main difficulty in PEPS is contraction, which is \#P-complete in
general~\cite{Schuch2007,Haferkamp2020}. Practical calculations therefore rely on approximate
contraction schemes such as the corner transfer matrix renormalization group
(CTMRG)~\cite{Nishino1996,Orus2009} and boundary-MPS methods~\cite{Vanderstraeten2022,Fishman2018}.
Even with these schemes, contraction cost and accuracy remain key limitations.

Isometric tensor network states (isoTNS)~\cite{Zaletel2020} form a PEPS-like two-dimensional ansatz
with local isometric constraints. 
The isometric constraints simplify contraction by defining an orthogonality column, 
which reduces local contractions to one-dimensional MPS contractions.
The column can be shifted by the Moses move~\cite{Zaletel2020}, with leading finite-size cost $O(D^7)$ for fixed physical dimension, where $D$ is the virtual bond dimension; 
its infinite version extends isoTNS to infinite strip geometries~\cite{Wu2023}.
The isometric structure also gives a natural connection to quantum circuits, 
since each local tensor can be embedded into a unitary gate acting on virtual and physical qubits~\cite{Slattery2021}. 
This correspondence has motivated quantum-circuit approaches to tensor-network ground-state optimization~\cite{Liu2019vqe,Leontica2025}.
It also gives a dynamical view of contraction: Malz and Trivedi~\cite{Malz2025}
formulated isoTNS contraction in terms of quantum-channel dynamics and analyzed
its computational complexity, and Dektor \textit{et al.}~\cite{Dektor2026}
introduced Born-rule sampling algorithms for finite two-dimensional isoTNS.

\begin{figure}[t!]
  \centering
  \includegraphics[width=\textwidth]{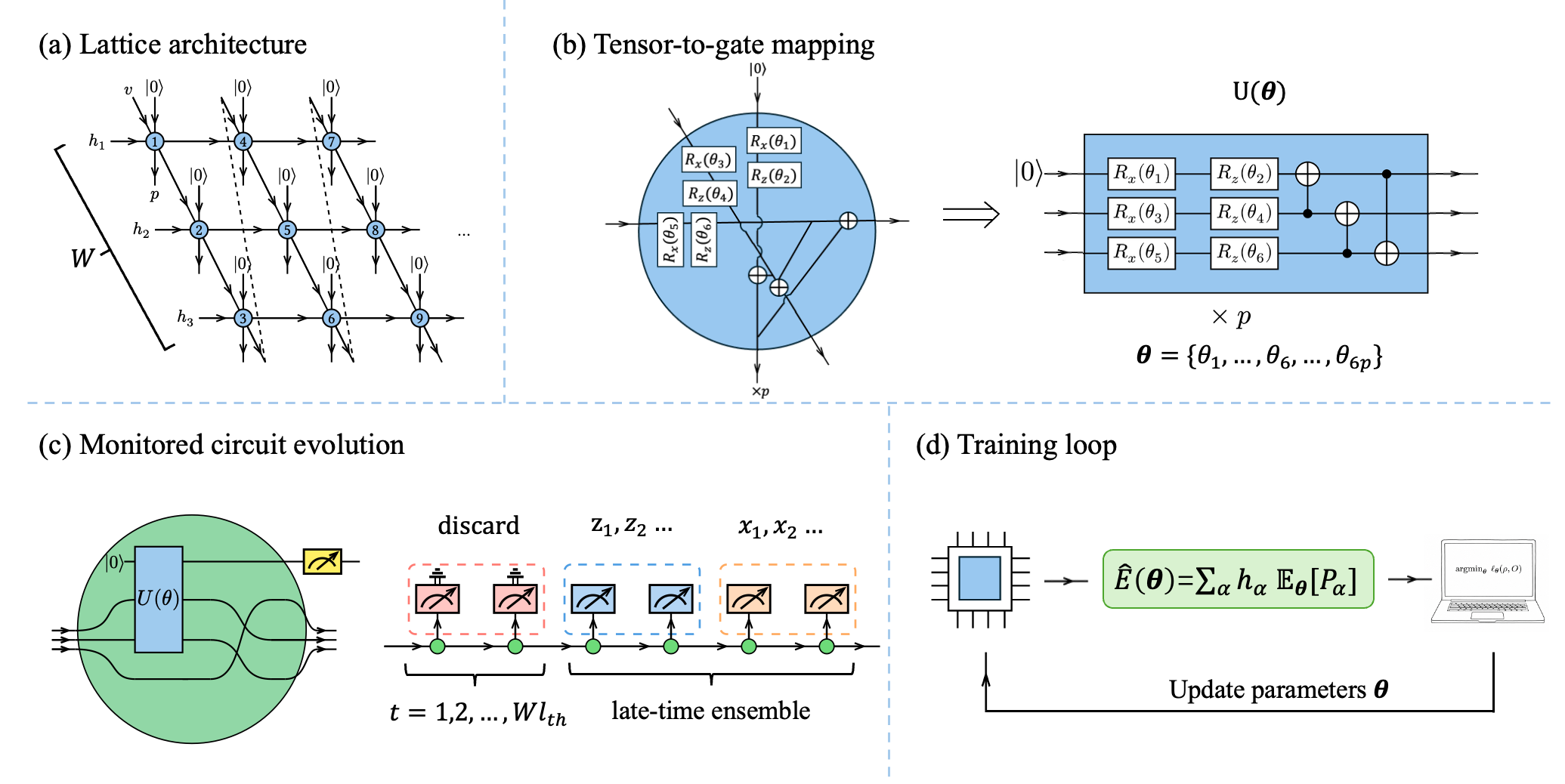}
  \caption{%
  \textbf{Variational sampling framework for isoPEPS.}
  (a)~An isoPEPS on an infinite cylinder, represented in a spiral geometry with $W$ sites around the circumference.
  Each site carries a rank-five PEPS tensor with four virtual legs---one incoming and one outgoing leg along each of the horizontal ($h_i$) and spiral ($v$) directions---and one physical leg $p$.
  The arrows specify the isometric orientation: the two incoming virtual indices are mapped to the physical index and the two outgoing virtual indices.
  The $\ket{0}$ line is an ancillary input introduced in the unitary circuit representation, rather than an additional leg of the PEPS tensor.
  (b)~The local isometric tensor is embedded in a three-qubit unitary $U(\vtheta)$, which is compiled into a sequence of single- and two-qubit unitaries. The right panel is a planar visualization of the same logical circuit as in the left panel. 
  The circuit ansatz consists of $p$ repetitions of single-qubit $R_x$ and $R_z$ rotations followed by a fixed controlled-NOT (CNOT) entangling pattern, yielding $\vtheta=\{\theta_1,\ldots,\theta_{6p}\}$.
  (c)
  The left panel shows a building block of the monitored circuit, corresponding to a local tensor of the isoPEPS, where the physical leg is projectively measured and reset such that this qubit can be reused for the next site. 
  The three virtual-qubit lines are grouped into the single thicker virtual bond shown on the right.
  The first $W l_{\mathrm{th}}$ measurement outcomes are discarded as thermalization; the late-time records are then collected in sequential basis blocks---first the $Z$ basis (outcomes $z_1, z_2, \ldots$), then the $X$ basis (outcomes $x_1, x_2, \ldots$)---and each Pauli term is estimated from the block measured in its matching basis, especially the energy terms supplied to the training loop.
   (d)~The estimated expectation values are combined with the Hamiltonian coefficients to form the sampled energy estimator $\widehat E(\vtheta)=\sum_\alpha h_\alpha \mathbb{E}_{\vtheta}[P_\alpha]$, which is passed to a classical optimizer that updates the variational parameters.
  } 
  \label{fig:pipeline}
\end{figure}

Building on these developments, we study the translationally invariant isometric PEPS (isoPEPS) geometry shown in Fig.~\ref{fig:pipeline}(a), an infinite cylindrical spiral.
Translation invariance along the infinite direction permits a fixed-point treatment: contracting successive columns is summarized by one repeated column transfer acting on the virtual boundary. 
The spiral geometry orders the column transfer sequentially, while the isometric constraints allow each local tensor map to be embedded in a unitary gate; together, these features yield a sequential quantum-circuit implementation.
We argue that the cylindrical spiral PEPS entails only a limited loss of generality: at the cost of an increased bond dimension, it can represent a conventional cylindrical PEPS (Appendix~\ref{app:spiral_equivalence}). 
Owing to the isometric constraints, this transfer matrix can also be interpreted as a completely positive trace-preserving (CPTP) boundary quantum channel~\cite{Malz2025}.
Its stationary state is the environment of the infinite cylindrical spiral.
Computing this stationary state exactly requires storing and evolving a boundary density matrix whose Hilbert-space dimension grows exponentially with the circumference $W$; the same transfer process, however, is naturally realized as a quantum circuit acting on a register of only $O(W\log_2 D)$ qubits.

The key insight is to unravel this boundary channel by measuring the physical qubits as each column is applied~\cite{Dalibard1992,Molmer1993,Plenio1998,Carmichael1993}.
If the measurement outcomes are discarded, one recovers the usual mixed-state channel evolution. 
If they are retained, the virtual boundary state follows a stochastic quantum trajectory~\cite{Plenio1998,Cheng2023}. 
This construction is related to monitored random circuits, where Born-rule measurement records can drive measurement-induced entanglement transitions~\cite{Bao2020,Jian2020}.
Here, however, the measurements unravel a fixed PEPS boundary channel, and the records are used to sample its stationary process rather than to tune across such a transition.
Under standard ergodicity assumptions, the time average of an infinite trajectory converges to the stationary boundary state~\cite{Kummerer2004}. 
Thus the measurement record both unravels the boundary channel and supplies samples for estimating physical observables,
without explicitly evolving the full boundary density matrix.
This distinguishes our approach from prior work in two ways: in contrast to the finite-size Born-rule sampling of Ref.~\cite{Dektor2026}, a single ergodic trajectory here samples the stationary state of the infinite system directly; and beyond the channel formulation of Ref.~\cite{Malz2025}, the sampled record serves as the noisy objective of a closed variational optimization loop.

We implement the variational ansatz by parametrizing each local isometry as a unitary gate composed of $p$ repeated blocks of single-qubit rotations and entangling gates.
Measuring and resetting physical qubits after their interaction with the virtual boundary allows the physical qubits to be reused during the iteration, 
reducing the active register size. 
Neutral-atom Rydberg arrays provide a concrete hardware route for these primitives; see Sec.~\ref{sec:discussion}. 

We apply the framework to two square-lattice spin models. 
As a validation test, we first consider the transverse-field Ising model (TFIM), where the optimized energy density across
different transverse fields agrees with DMRG and other numerical references~\cite{HibatAllah2020,Lubasch2014}.
We then study the frustrated square-lattice $J_1$--$J_2$ Heisenberg model, whose phase behavior involves competing N\'eel, stripe, and debated intermediate nonmagnetic regimes, including spin-liquid and valence-bond solid (VBS) scenarios~\cite{Jiang2012,Gong2014,Nomura2021}.
The magnetic and dimer order parameters in Fig.~\ref{fig:j1j2_m2} distinguish these regimes, and the momentum-resolved spin and dimer structure factors in Fig.~\ref{fig:j1j2_structure_factors} are consistent with them.
Together, these results demonstrate that monitored isoPEPS provide a reliable framework for variational studies of two-dimensional quantum phases.


\section{Method}
\label{sec:method}

\subsection{isoPEPS as parameterized quantum circuits}
\label{sec:method_circuit}
We consider a two-dimensional quantum spin system on the infinite, translationally invariant cylindrical spiral of circumference $W$ shown in Fig.~\ref{fig:pipeline}(a).
The equivalence between this spiral representation and an ordinary cylindrical PEPS is established in Appendix~\ref{app:spiral_equivalence}.
Each site of the isoTNS~\cite{Zaletel2020} is an isometric tensor $A$, mapping the two incoming virtual legs to the physical leg and the two outgoing virtual legs, subject to the isometry condition:
\begin{equation}
\rankSixTensorIcon\;:\; \mathbb{C}^{D^2} \to \mathbb{C}^d \otimes \mathbb{C}^{D^2}, \quad A^\dagger A = I,
\label{eq:isometry}
\end{equation}
where $d$ is the physical dimension and $D$ is the virtual bond dimension.
The isometric constraint ensures an exact mapping to a unitary gate, which can be viewed as a rank-six tensor after completing the isometry.

We can complete the remaining orthogonal basis of $A$ to obtain a unitary gate $U$, which reproduces the isometry when its physical input is fixed to $\ket{0}$:
\begin{equation}
U\left(\ket{0}_{\mathrm{phys}} \otimes \ket{\psi}_{\mathrm{virt}}\right)
= A\ket{\psi}_{\mathrm{virt}}
= \sum_{s} \ket{s}_{\mathrm{phys}} \otimes K_s \ket{\psi}_{\mathrm{virt}},
\qquad
K_s \equiv \bra{s}_{\mathrm{phys}} U \ket{0}_{\mathrm{phys}} ,
\label{eq:gate_map}
\end{equation}
where $\ket{s}$ runs over the measurement basis of the physical qubit and the operators $K_s$ act on the virtual factor.
The isometry condition implies $\sum_s K_s^\dagger K_s = A^\dagger A = I$, so $\{K_s\}$ forms a valid set of Kraus operators; these generate the measurement unraveling of Sec.~\ref{sec:method_unraveling}.
The gate acts on $n_q = \log_2 d + 2\log_2 D$ qubits---one physical factor and two (vertical and horizontal) virtual factors---so $U\in SU(2^{n_q})$.

We parametrize each gate $U$ as a product of $p$ layers, where each layer consists of single-qubit rotations $R_x(\theta) R_z(\theta)$ on every qubit 
followed by a fixed entangling pattern of CNOT gates to generate entanglement between qubits.
This gives $2 p n_q$ real parameters per gate. The resulting parameterized gate is denoted $U(\vtheta)$ and is shown in Fig.~\ref{fig:pipeline}(b).

As illustrated in Fig.~\ref{fig:pipeline}(a), columns are linked sequentially via the vertical virtual qubits,
so that the output vertical virtual state of one column becomes the input for the next,
forming a sequential quantum circuit along the cylindrical spiral. The total number of qubits needed is $n_t = \log_2 d + (W+1) \log_2 D$.
Translational invariance along the infinite direction allows the same gates to be reused: the ansatz repeats a unit cell of a few independent gates, whose period is chosen to accommodate the orders expected in the model at hand (specified for each model in Sec.~\ref{sec:results}).
This choice balances expressivity with computational efficiency, allowing the circuit to capture ground-state correlations while keeping the parameter count manageable.

\subsection{Unravelling the quantum channel of transfer matrix}
\label{sec:method_unraveling}


For an isoPEPS on an infinite cylindrical spiral, one column of tensors defines a transfer matrix on the virtual boundary.
Because the local tensors are isometric, the transfer matrix can be written as a CPTP map
\begin{equation}
\rho_l \;\longrightarrow\; \underbrace{\threeTensorColumnMap}_{\mathcal{E}} \;\longrightarrow\; \rho_{l+1},
\end{equation}
where $\rho_l$ is the boundary density matrix after $l$ columns. If the channel is mixing, repeated application of $\mathcal{E}$ converges to a stationary state
\begin{equation}
\mathcal{E}(\rho^\ast)=\rho^\ast .
\end{equation}

The repeated channel admits a monitored-circuit unraveling. Instead of tracing out the physical qubits in each column, we measure them and keep the outcomes. 
A single measurement record then defines a conditional quantum trajectory of the boundary state, while averaging over all possible records recovers the
unmonitored channel evolution. Thus exact contraction and monitored sampling are two representations of the same transfer process.

Concretely, one application of $\mathcal{E}$ corresponds to sweeping through a column of circumference $W$. At site $j$, we apply the local unitary, measure the
physical qubit with outcome $s_j$, reset it to $\ket{0}$, and pass the updated virtual state to the next site. Let $K^{(j)}_{s_j}$ be the Kraus operator
of Eq.~\eqref{eq:gate_map} associated with measurement outcome $s_j$ at site $j$, acting on the corresponding virtual factors, as shown in Fig.~\ref{fig:pipeline}(c). Starting from $\rho^{(0)}=\rho_0$, the conditional probability for outcome $s_j$ is
\begin{equation}
p(s_j|s_1,\ldots,s_{j-1})
=
\tr\!\left[
K^{(j)}_{s_j}\rho^{(j-1)}K^{(j)\dagger}_{s_j}
\right],
\end{equation}
and the normalized conditional boundary state is
\begin{equation}
\rho^{(j)}
=
\frac{
K^{(j)}_{s_j}\rho^{(j-1)}K^{(j)\dagger}_{s_j}
}{
\tr\!\left[
K^{(j)}_{s_j}\rho^{(j-1)}K^{(j)\dagger}_{s_j}
\right]
}.
\end{equation}

After a complete column sweep, the ordered outcomes form the record $\mathbf{s}=(s_1,\ldots,s_W)$, with effective Kraus operator
\begin{equation}
K_{\mathbf{s}}
=
K^{(W)}_{s_W}\cdots K^{(1)}_{s_1}.
\end{equation}
The probability of the full record is
\begin{equation}
p(\mathbf{s}|\rho)
=
\tr\!\left[
K_{\mathbf{s}}\rho_0 K_{\mathbf{s}}^\dagger
\right],
\end{equation}
and summing over all records gives the column channel
\begin{equation}
\mathcal{E}(\rho_0)
=
\sum_{\mathbf{s}}K_{\mathbf{s}}\rho_0 K_{\mathbf{s}}^\dagger,
\qquad
\sum_{\mathbf{s}}K_{\mathbf{s}}^\dagger K_{\mathbf{s}}=I .
\end{equation}

For a system with $L$ columns,
$\Gamma_L=(\mathbf{s}^{(1)},\ldots,\mathbf{s}^{(L)})$, the ensemble average over all trajectories reproduces the deterministic channel evolution,
\begin{equation}
\sum_{\Gamma_L}p(\Gamma_L|\rho_0)\,\rho_L^{(\Gamma_L)}
=
\mathcal{E}^L(\rho_0).
\end{equation}

A single long monitored trajectory can sample the stationary process generated by $\rho^\ast$. Under the usual ergodicity
assumption for the induced quantum trajectory,
\begin{equation}
\frac{1}{L}\sum_{l=1}^{L}\rho_l^{(\Gamma)}
\xrightarrow[L\to\infty]{\mathrm{a.s.}}
\rho^\ast .
\end{equation}

Therefore, for an observable $O$ spanning $c$ consecutive columns, let
$o(\mathbf{s}^{(l)},\ldots,\mathbf{s}^{(l+c-1)})$ denote its estimator
constructed from the corresponding measurement records. Averaging these
estimator values along a single trajectory gives

\begin{equation}
\widehat O_L
=
\frac{1}{L}\sum_{l=1}^{L}
o\!\left(
\mathbf{s}^{(l)},\ldots,\mathbf{s}^{(l+c-1)}
\right),
\end{equation}
with
\begin{equation}
\widehat O_L
\xrightarrow[L\to\infty]{\mathrm{a.s.}}
\langle O\rangle_{\rho^\ast}.
\end{equation}
Here $\mathrm{a.s.}$ denotes almost-sure convergence.
This is the trajectory-unravelling analogue of Monte Carlo sampling: a single typical monitored path samples the stationary quantum process, and time averages along
that path reproduce equilibrium expectation values.

In practical simulations, the monitored trajectory has finite length, so the stationary regime can only be reached approximately. 
To reduce the bias from the initial transient, we discard the first $l_{\mathrm{th}}$ columns as thermalization and estimate observables from the remaining $L_{\mathrm{s}}$ columns,
\begin{equation}
\widehat O_{L_{\mathrm{s}}}
=
\frac{1}{L_{\mathrm{s}}}
\sum_{l=l_{\mathrm{th}}+1}^{l_{\mathrm{th}}+L_{\mathrm{s}}}
o\!\left(
\mathbf{s}^{(l)},\ldots,\mathbf{s}^{(l+c-1)}
\right).
\end{equation}
Each observable term is measured in the eigenbasis of its Pauli components---$X$, $Y$, or $Z$.
Measuring in different bases corresponds to different Kraus decompositions in Eq.~\eqref{eq:gate_map} and hence to different unravelings of the same channel $\mathcal{E}$, whose ensemble average is basis independent.
We therefore divide the post-thermalization portion of a single trajectory into sequential blocks, one per required basis---first $Z$, then $X$ (and $Y$ where needed)---and estimate each Pauli term from the records of the block measured in its matching basis, as illustrated in Fig.~\ref{fig:pipeline}(c).
Because every unraveling averages to the same channel, switching the measurement basis between blocks does not disturb the stationary statistics being sampled.

The convergence rate toward stationarity is governed by the spectral gap~$\Delta$ of the transfer matrix. 
For a gapped system, deviations from the stationary state decay as
\begin{equation}
\|\rho_l-\rho^\ast\|_1 \leq C e^{-l\Delta},
\end{equation}
so the thermalization length scales as $l_{\mathrm{th}}=O(1/\Delta)$. 
We diagnose this convergence by tracking representative local observables, such as the local energy, 
as functions of the iteration step~$l$, where one iteration step corresponds to adding one column. The relaxation diagnostic in Fig.~\ref{fig:tfim_energy} shows this behavior.
In this diagnostic, all trajectories are initialized from the same state, and the value at fixed~$l$ is obtained by averaging the observable over these
independent trajectories. This average can be interpreted as the expectation value of the observable in the ensemble-averaged mixed state~$\rho_l$.
The relaxation of these observables toward a stable plateau indicates that the ensemble has entered the stationary regime, and determines
how many initial columns should be discarded before performing time averages along a long trajectory.
For both TFIM and \(J_1\)–\(J_2\) Heisenberg model, we discard the first \(l_{\mathrm{th}}=100\) steps. 
Fig.~\ref{fig:tfim_energy} shows that the measured energy has already reached its stable plateau by \(l_{\mathrm{th}}\approx10\) in the representative TFIM case; we nevertheless use \(l_{\mathrm{th}}=100\) throughout as a conservative choice.

\subsection{Variational optimization}

The circuit parameters~$\vtheta$ are optimized through the hybrid quantum-classical training loop shown in Fig.~\ref{fig:pipeline}(d).
For each candidate parameter vector, we generate a monitored trajectory from the corresponding circuit, discard the first~$l_{\mathrm{th}}$ columns to
remove the initial transient, and estimate the energy from the remaining measurement records.

For a Hamiltonian decomposed into local Pauli terms,
\begin{equation}
H=\sum_{\alpha} h_{\alpha} P_{\alpha},
\end{equation}
where $P_{\alpha}$ are Pauli-string operators and $h_{\alpha}$ are their real coefficients.
The sampled energy is estimated as
\begin{equation}
\widehat{E}(\vtheta)
=
\sum_{\alpha} h_{\alpha}\,
\mathbb{E}_{\vtheta}[P_{\alpha}],
\end{equation}
where $\mathbb{E}_{\vtheta}[P_{\alpha}]$ denotes the average of the corresponding Pauli-string measurement over the retained samples.

Because each sampled energy evaluation is noisy, we optimize the variational parameters using the covariance matrix adaptation evolution strategy (CMA-ES)~\cite{Hansen2006},
a derivative-free method well suited to noisy objective functions. 
Each candidate parameter vector is evaluated by sampling a monitored trajectory, and the resulting sampled
energies are used to update the parameters for the next generation.
For all optimizations, CMA-ES uses population size \(\lambda=4+\lfloor3\ln N_\theta\rfloor\), where \(N_\theta\) is the number of variational parameters, initial step size \(\sigma_0=0.1\), and at most \(1000\) generations. 
Parameters are initialized uniformly in \([0,1)\) with Random.seed!(123) and constrained to \([0,2\pi)\) during optimization. 
After optimization, we retain the \(20\) candidates with the lowest sampled energies and select the final parameters with lowest energy by exact tensor-network contraction.


\section{Results}
\label{sec:results}

\subsection{TFIM benchmark}
As a validation test, we apply this method to the two-dimensional TFIM
\begin{equation}
H = -J \sum_{\langle ij\rangle} Z_i Z_j - g \sum_i X_i ,
\end{equation}
where $\langle ij\rangle$ denotes nearest-neighbor pairs on the cylinder, we set $J=1$, and $g$ is the transverse-field strength.
We optimize a $1\times3$ isoPEPS ansatz with bond dimension $D=2$ and $p=3$ repeated circuit blocks on a cylindrical spiral of circumference $W=3$ using monitored-circuit sampling with $60,000$ samples.
The $1\times3$ unit cell repeats three independent gates with A-B-C periodicity: although the TFIM Hamiltonian is invariant under translations around the cylinder, the spiral geometry distinguishes the rows within a column---in particular the first row carries the enlarged bond of the spiral construction (Appendix~\ref{app:spiral_equivalence})---so we allow one independent gate per row; the uniform A-A-A ansatz is compared in Table~\ref{tab:tfim_literature}.
Here one sample denotes one retained post-thermalization single-site measurement outcome of the monitored trajectory, cf.\ Fig.~\ref{fig:pipeline}(c).
We then compute the energy density of each optimized state by exact tensor-network contraction. 
We emphasize that sampling and evaluation play distinct roles here: the variational parameters are optimized from noisy energies estimated along the monitored measurement records, whereas the benchmark energy densities reported in this section are obtained by exact tensor-network contraction of the resulting optimized state. This comparison therefore validates the sampling-based optimization, while the accuracy of the sampled estimator itself is characterized in Appendix~\ref{app:tfim}.
Figure~\ref{fig:tfim_energy_vs_g} compares our energy densities with DMRG
results on a $3\times1000$ cylinder, obtained in this work using
ITensors.jl~\cite{ITensor} with the energy density extracted from the central
bulk region. Representative literature benchmarks are
summarized in Table~\ref{tab:tfim_literature}.
Relative to the same-geometry VUMPS reference in Table~\ref{tab:tfim_literature}, the optimized energy densities deviate by $0.4\%$ at $g=2.0$, $2.2\%$ at $g=3.0$, and $0.9\%$ at $g=4.0$; the largest deviation occurs near the critical region, as expected for the modest bond dimension $D=2$.
This level of agreement supports the reliability of the sampling-based optimization. 
Additional TFIM diagnostics in Appendix~\ref{app:tfim} quantify thermalization, finite-sample convergence, correlation growth, magnetization, and sensitivity to readout errors.
\begin{table}[t]
\caption{\textbf{Representative energy benchmarks for the TFIM.}
All entries are quoted as energy densities for
$H=-J\sum_{\langle ij\rangle}Z_iZ_j-g\sum_iX_i$ with $J=1$.
Open boundary conditions (OBC) are noted where applicable.
Because the quoted results correspond to distinct geometries and boundary conditions, the table is intended as a contextual benchmark rather than a direct comparison.
}
\label{tab:tfim_literature}
\begingroup
\normalsize
\setlength{\tabcolsep}{3pt}
\renewcommand{\arraystretch}{0.92}
\begin{tabular}{>{\raggedright\arraybackslash}p{0.34\textwidth} >{\raggedright\arraybackslash}p{0.16\textwidth} rrr}
\toprule
Method & Geometry & \multicolumn{3}{c}{$E_0/N_{\mathrm{site}}$} \\
 & & $g=2.0$ & $g=3.0$ & $g=4.0$ \\
\midrule
finite PEPS, full update (FU), $D=4$~\cite{Lubasch2014} & $21\times21$ OBC & $-2.45219(1)$ & $-3.18243(1)$ & $-4.12694(1)$ \\
DMRG values as reported in Ref.~\cite{HibatAllah2020} & $12\times12$ OBC & $-2.40960263$ & $-3.17389966$ & $-4.12179793$ \\
variational uniform matrix product state (VUMPS), $D=32$ (this work) & $3 \times \infty$ cylinder, A-B-C & $-2.512022$ & $-3.217557$ & $-4.144629$ \\
isoPEPS (this work) & $3 \times \infty$ cylinder, A-B-C & $-2.501703$ & $-3.148047$ & $-4.106446$ \\
isoPEPS (this work) & $3 \times \infty$ cylinder, A-A-A & $-2.500258$ & $-3.139882$ & $-4.087263$ \\
\bottomrule
\end{tabular}
\endgroup
\begin{minipage}{\textwidth}
\footnotesize
Here OBC marks open-boundary geometries.
The labels A-B-C and A-A-A denote, respectively, the three-independent-gate ansatz and the uniform single-gate ansatz (all gates in the unit cell identical).
The VUMPS values were obtained in this work using MPSKit.jl~\cite{MPSKit}.
\end{minipage}
\end{table}

\begin{inlinefigurecontent}
  \subfigure[]{\includegraphics[width=0.49\columnwidth]{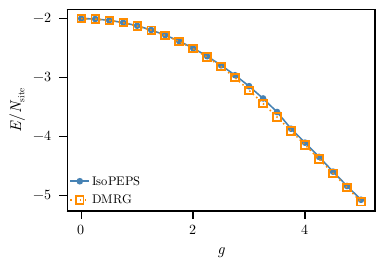}}
  \hfill
  \subfigure[]{\includegraphics[width=0.49\columnwidth]{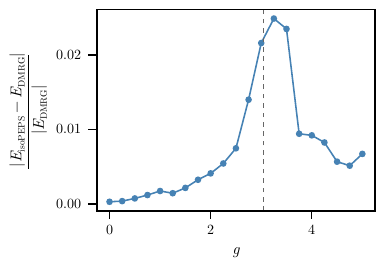}}
  \caption{\textbf{TFIM energy benchmark on a $3\times\infty$ cylinder.}
(a)~Energy density of the $D=2$, $p=3$ isoPEPS after sampling-based optimization, evaluated by exact tensor-network contraction, compared with DMRG on the same $W=3$ cylinder. 
(b)~Relative deviation of the exact-contraction isoPEPS energy from the DMRG result,
$|E_{\mathrm{isoPEPS}}-E_{\mathrm{DMRG}}|/|E_{\mathrm{DMRG}}|$.
The deviation peaks near the square-lattice quantum critical point $g_c=3.04438(2)$~\cite{Blote2002}. 
}
  \label{fig:tfim_energy_vs_g}
\end{inlinefigurecontent}

\subsection[Frustrated J1--J2 Heisenberg model]{Frustrated $J_1$--$J_2$ Heisenberg model}
\label{sec:j1j2}
We next apply the framework to the frustrated spin-$1/2$ Heisenberg $J_1$--$J_2$ model on the square-lattice. The Hamiltonian is
\begin{equation}
H = J_1 \sum_{\langle ij\rangle} \mathbf{S}_i \cdot \mathbf{S}_j
  + J_2 \sum_{\langle\!\langle ij\rangle\!\rangle} \mathbf{S}_i \cdot \mathbf{S}_j ,
\end{equation}
where $\langle ij\rangle$ and $\langle\!\langle ij\rangle\!\rangle$ denote nearest-neighbor and next-nearest-neighbor pairs, respectively, with $\mathbf{S}_i \cdot \mathbf{S}_j = \frac{1}{4}(X_i X_j + Y_i Y_j + Z_i Z_j)$.
We set $J_1=1$ and vary $J_2/J_1 \in [0,1]$. The model exhibits N\'eel antiferromagnetism at small $J_2/J_1$ and stripe order at larger $J_2/J_1$, while the intermediate frustrated region has a debated ground-state character, with spin-liquid and valence-bond-ordered scenarios proposed across different numerical approaches~\cite{Dagotto1989,Haghshenas2018,Jiang2012,Gong2014,Liu2022,Qian2024}.

To allow the competing N\'eel, stripe, and period-two bond patterns within a common ansatz, 
we use a $2\times2$ unit cell with four independent gates on a cylindrical spiral of circumference $W=4$.
The ansatz uses $p=3$ repeated circuit blocks per unitary gate and bond dimension $D=2$, corresponding to $n_q=3$ qubits per gate and $72$ variational parameters in total.


Figure~\ref{fig:j1j2_opt_convergence} shows the training history energy
at $J_2/J_1=0.5$. At each training step, the energy density is estimated
from $N_{\mathrm{shot}}=40{,}000$ measurement samples generated by the monitored
circuit. The sampled energy decreases during approximately $200$
steps and then approaches a plateau. The dashed horizontal lines mark the
reference result from DMRG with $D=192$ and $D=2$. The subsequent fluctuations around this value are
consistent with the stochastic objective used by CMA-ES and indicate that
the optimization has reached a stable energy within the finite-sampling
resolution. 
For the full $J_2/J_1$ scan, we discard the first $l_{\mathrm{th}}=100$ channel iterations.
Quantitative accuracy is assessed in Appendix~\ref{app:j1j2}: At $J_2/J_1=0.5$, the $D=2$ isoPEPS energy is $0.02$ per site above
the $D=2$ DMRG result and $0.12$ per site above the $D=192$ DMRG result. 
Across the scan, the isoPEPS energy follows the $D=2$ DMRG reference but remains above the more converged $D=192$ result, with the largest deviation in the intermediate frustrated regime.

\begin{inlinefigurecontent}
  \subfigure[]{\includegraphics[width=0.7\columnwidth]{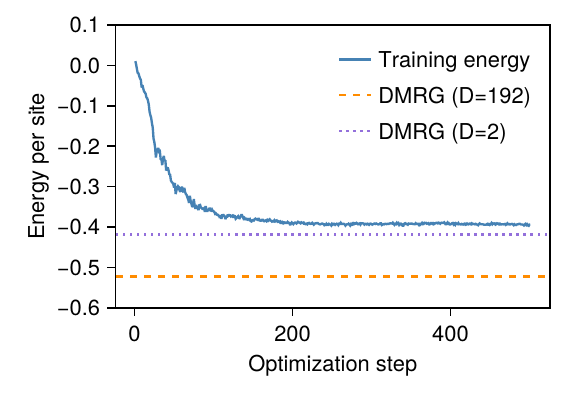}\label{fig:j1j2_opt}}
  \caption{%
  \textbf{Optimization history for the $J_1$-$J_2$ Heisenberg model at $J_2/J_1=0.5$.}
  IsoPEPS with $W=4$, $D=2$, $p=3$, and a $2\times2$ unit cell ($72$ parameters).
  Each energy estimate uses $40{,}000$ single-site measurement samples from the monitored trajectory.
  %
  }
  
  \label{fig:j1j2_opt_convergence}
\end{inlinefigurecontent}

\paragraph{Magnetic and dimer correlations.}

To distinguish competing magnetic and dimer orders, we evaluate the spin and connected dimer structure factors
\begin{equation}
S(\mathbf q) = \frac{1}{N_{\rm site}}\!\sum_{i,j} e^{i\mathbf q\cdot(\mathbf r_i-\mathbf r_j)} \langle \mathbf S_i\cdot \mathbf S_j\rangle ,
\qquad
S_D^\mu(\mathbf q) = \frac{1}{N_b}\!\sum_{i,j} e^{i\mathbf q\cdot(\mathbf r_i-\mathbf r_j)} \langle \delta D_i^\mu\, \delta D_j^\mu\rangle ,
\end{equation}
where the index $\mu\in{x,y}$ labels the two axes of the square lattice and $\hat\mu$ is the corresponding unit vector, $D_i^\mu = \mathbf S_i\cdot \mathbf S_{i+\hat\mu}$ is the nearest-neighbor bond operator along direction $\mu$, $\delta D_i^\mu = D_i^\mu - \langle D_i^\mu\rangle$, and $N_{\rm site}$ and $N_b$ are the numbers of lattice sites and bonds of the chosen orientation.  

The normalized order parameters $M^2(\mathbf q)=S(\mathbf q)/N_{\rm site}$ and $M_D^2(\mathbf q)=S_D^y(\mathbf q)/N_b$ are evaluated at the characteristic wavevectors: $M^2(\pi,\pi)$ probes N\'eel order, $M^2(0,\pi)$ probes stripe order, and $M_D^2(0,\pi)$ probes columnar VBS order.
The choice of dimer observable is guided by the real-space correlations: the dimer pattern occurs on vertical ($y$-oriented) bonds and alternates along the $y$ direction.  We therefore analyze the vertical-bond dimer structure factor $S_D^y(\mathbf q)$ and use its normalized value $M_D^2(0,\pi)$ as the dimer order parameter.  In contrast, $M^2(0,\pi)$ is the magnetic order parameter used to detect stripe order.  The alternatives $S_D^x(\mathbf q)$ and $M_D^2(\pi,0)$ would probe a horizontal-bond pattern or modulation along $x$, neither of which is observed.  This directional selection is natural on the $W=4$ cylinder, where the $x$ and $y$ bond directions are inequivalent and the geometry can pin the columnar orientation.
The corresponding real-space bond-correlation patterns are shown in Fig.~\ref{fig:j1j2_bond_correlations} of Appendix~\ref{app:j1j2}.
As in the TFIM benchmark, the variational parameters are optimized from sampled energies, whereas the correlators entering $S(\mathbf q)$ and $S_D^y(\mathbf q)$ are evaluated by exact tensor-network contraction of the optimized state over $21$ consecutive columns, exploiting translational invariance.
Since the peak height of a structure factor grows with the number of columns included, these quantities serve to identify the ordering wavevectors rather than as converged order-parameter estimates.

Figure~\ref{fig:j1j2_m2} traces the three order parameters across $J_2/J_1$ and distinguishes three regimes. At small $J_2/J_1$, $M^2(\pi,\pi)$ dominates. In the intermediate window $J_2/J_1\approx 0.5$--$0.6$, both magnetic components drop sharply while $M_D^2(0,\pi)$ peaks, identifying a columnar VBS regime. At larger $J_2/J_1$, $M^2(0,\pi)$ dominates and the dimer signal is suppressed. The isoPEPS results qualitatively agree with the DMRG calculations performed in this work: both show the suppression of N\'eel order, an enhanced dimer signal in the intermediate window, and the onset of stripe order at larger $J_2/J_1$, although their amplitudes do not coincide point by point.
Figure~\ref{fig:j1j2_structure_factors} shows the same evolution in momentum space at $J_2/J_1=0.0,\,0.5,\,0.6,\,1.0$. For visualization, the Fourier sums are evaluated on a dense momentum grid, although the $W=4$ circumference allows only $q_y=0,\,\pi/2,\,\pi,\,3\pi/2$ as transverse crystal momenta. The spin structure factor $S(\mathbf q)$ is concentrated at $(\pi,\pi)$ for $J_2/J_1=0$, is suppressed through the intermediate window, and shifts to $(0,\pi)$ at $J_2/J_1=1$. The dimer structure factor $S_D^y(\mathbf q)$ peaks at $(0,\pi)$ for $J_2/J_1\approx 0.5$--$0.6$ and is featureless outside this window, consistent with Fig.~\ref{fig:j1j2_m2}.

The $J_1$--$J_2$ model tests the framework in a frustrated regime where physical consistency across competing correlations is essential, and the momentum-resolved magnetic and dimer structure factors (Fig.~\ref{fig:j1j2_structure_factors}) together with the three order parameters (Fig.~\ref{fig:j1j2_m2}) show that the framework remains reliable beyond simple models like the TFIM.

\begin{inlinefigurecontent}
  \includegraphics[width=0.7\columnwidth]{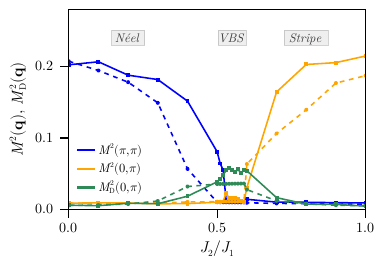}
  \caption{%
  \textbf{Magnetic and dimer order calculated for the $J_1$-$J_2$ Heisenberg model.}
  Magnetic order parameters at the N\'eel wavevector $M^2(\pi,\pi)$ and the stripe wavevector $M^2(0,\pi)$, together with the vertical-bond dimer order parameter $M_D^2(0,\pi)$, as a function of $J_2/J_1$.
  Colors denote the three observables; solid (dashed) curves show the $D=2$ isoPEPS (DMRG) results.
  The shaded labels mark the N\'eel, VBS, and stripe regimes: $M^2(\pi,\pi)$ dominates at small $J_2/J_1$, $M_D^2(0,\pi)$ peaks in the intermediate window $J_2/J_1\approx 0.5$--$0.6$, and $M^2(0,\pi)$ takes over at larger $J_2/J_1$.
  %
  %
  }
  \label{fig:j1j2_m2}
\end{inlinefigurecontent}

\begin{inlinefigurecontent}
    \includegraphics[width=0.9\linewidth]{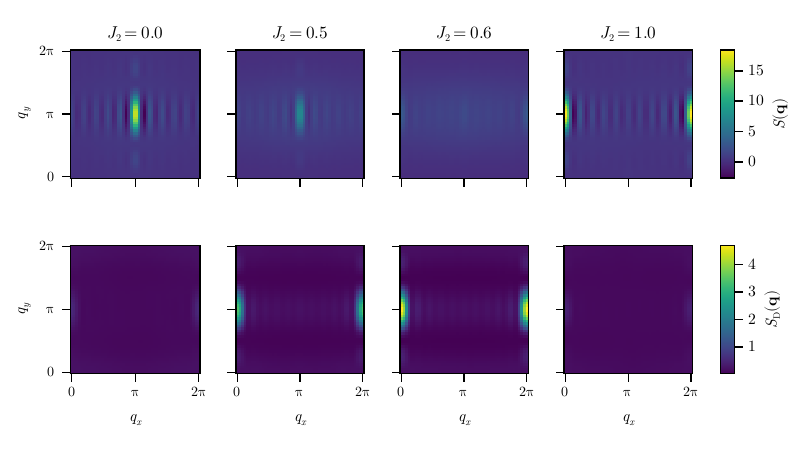}%
  \caption{%
  \textbf{Momentum-resolved spin and dimer structure factors for the $J_1$--$J_2$ Heisenberg model.}
  (Top row)~Spin structure factor $S(\mathbf q)$ at $J_2/J_1=0.0,\,0.5,\,0.6,\,1.0$: a N\'eel peak at $(\pi,\pi)$ for $J_2/J_1=0.0$, suppressed in the intermediate window, and shifted to the stripe wavevector $(0,\pi)$ at $J_2/J_1=1.0$.
  (Bottom row)~Vertical-bond dimer structure factor $S_D^y(\mathbf q)$ at the same four couplings: a peak at $(0,\pi)$ for $J_2/J_1=0.5,\,0.6$, consistent with the columnar-VBS signal in Fig.~\ref{fig:j1j2_m2}, and featureless at $J_2/J_1=0.0,\,1.0$.
  The dense $q_y$ sampling in the plots is used only for display; the $W=4$ cylinder admits the four transverse crystal momenta $q_y\in\{0,\,\pi/2,\,\pi,\,3\pi/2\}$.
  }
  \label{fig:j1j2_structure_factors}
\end{inlinefigurecontent}

\section{Discussion}
\label{sec:discussion}

We present a variational framework for studying two-dimensional quantum phases with isoPEPS on an infinite cylindrical spiral, in which one column of the tensor network defines a quantum channel on the virtual boundary degrees of freedom.
The framework builds on the isoTNS ansatz~\cite{Zaletel2020,Wu2023} and its quantum-channel interpretation~\cite{Malz2025}; its central idea is to unravel this channel into monitored quantum trajectories, so that after thermalization a single long trajectory samples the stationary boundary state, and the same measurement record yields Monte Carlo estimators for observables.
For the TFIM, the energy densities of the sample-optimized states agree with DMRG benchmarks and results from the literature.
For the frustrated $J_1$--$J_2$ Heisenberg model, the method captures the crossover from N\'eel to stripe correlations
and a vertical-dimer signal in the intermediate regime.

A natural next step is gradient-based optimization for this ansatz.
We use CMA-ES here because it copes well with noisy sampled objectives, but its cost grows quickly with the number of parameters.
Gradients obtained from the fixed-point channel or from trajectory-based estimators could allow the method to scale to larger bond dimensions and wider circumferences.

The framework should also extend to other lattice geometries, with the kagome lattice as an immediate target~\cite{Yan2011}.
Gapped topologically ordered phases~\cite{Kitaev2006,LevinWen2006} are another promising direction, since the measurement record may offer diagnostics that complement standard order parameters and entanglement-based probes.

Neutral-atom Rydberg arrays are a concrete platform for monitored trajectories: optical tweezer arrays provide programmable two-dimensional geometries and Rydberg-mediated entangling gates~\cite{Browaeys2020,Ebadi2021}, while recent neutral-atom processors have demonstrated mid-circuit readout, feed-forward, and reconfigurable circuit operations~\cite{Graham2023MidCircuit,Bluvstein2024}.
The practical performance will depend on the overhead of repeated measurements and resets and on the number of samples required to estimate observables accurately;
a simple independent-bit-flip model of readout errors and its effect on the TFIM energy estimator are examined in Figs.~\ref{fig:tfim_readout_energy} and~\ref{fig:tfim_readout_energy_error} of Appendix~\ref{app:tfim_readout}.
In such a hybrid implementation, quantum hardware generates the monitored trajectories while a classical optimizer updates the variational parameters, providing a direct path from the present simulations to experiment.

\acknowledgments{
  Jinguo Liu acknowledges the support of the National Natural Science Foundation of China (grant no.~12404568).
  Guo-Yi Zhu acknowledges the support of National Natural Science Foundation of China - Young Scientists Fund (grant no.~12504181), Start-up Fund of the Hong Kong University of Science and Technology (Guangzhou) (HKUST(GZ)) (grant no.~G0101000221), Guangdong provincial project (grant no.~2024QN11X201) and Guangdong Basic and Applied Basic Research Foundation (grant no.~2026A1515010965).
}

\section*{Contribution Statement}
Yuqing Rong implemented the numerical methods and performed the numerical simulations. Huan-Hai Zhou carried out theoretical validation and cross-checked the numerical results. Jinguo Liu proposed the core methodology and supervised the project. Guo-Yi Zhu supervised the project and provided scientific guidance. All authors discussed the results and contributed to the revision of the manuscript.

\section*{Data and Code Availability}
The numerical data supporting the findings of this study are openly available on Zenodo~\cite{Rong2026IsoPEPSData}. The source code used to generate and analyze the results is available in the \texttt{IsoPEPS.jl} v1.0.0 release at \url{https://github.com/yuqingrong/IsoPEPS.jl/releases/tag/v1.0.0}.

\bibliographystyle{quantum}
\bibliography{groupbib}

@article{Zaletel2020,
  title={Isometric Tensor Network States in Two Dimensions},
  author={Zaletel, Michael P. and Pollmann, Frank},
  journal={Physical Review Letters},
  volume={124},
  pages={037201},
  year={2020},
  doi={10.1103/PhysRevLett.124.037201}
}

@article{Wu2023,
  title={Two-dimensional isometric tensor networks on an infinite strip},
  author={Wu, Yantao and Anand, Sajant and Lin, Sheng-Hsuan and Pollmann, Frank and Zaletel, Michael P.},
  journal={Physical Review B},
  volume={107},
  pages={245118},
  year={2023},
  doi={10.1103/PhysRevB.107.245118}
}

@article{Malz2025,
  title={Computational Complexity of Isometric Tensor-Network States},
  author={Malz, Daniel and Trivedi, Rahul},
  journal={PRX Quantum},
  volume={6},
  pages={020310},
  year={2025},
  doi={10.1103/PRXQuantum.6.020310}
}

@misc{Slattery2021,
  title={Quantum circuits for two-dimensional isometric tensor networks},
  author={Slattery, Samuel A. and Clark, Bryan K.},
  year={2021},
  eprint={2108.02792},
  archivePrefix={arXiv},
  primaryClass={quant-ph}
}

@misc{Dektor2026,
  title={Sampling two-dimensional isometric tensor network states},
  author={Dektor, Alec and Dumitrescu, Eugene F. and Yang, Chao},
  year={2026},
  eprint={2602.02245},
  archivePrefix={arXiv},
  primaryClass={quant-ph}
}

@misc{Leontica2025,
  title={Optimizing two-dimensional isometric tensor networks with quantum computers},
  author={Leontica, Sebastian and Baiardi, Alberto and Schuhmacher, Julian and Tacchino, Francesco and Tavernelli, Ivano},
  year={2025},
  eprint={2511.13827},
  archivePrefix={arXiv},
  primaryClass={quant-ph}
}

@misc{Verstraete2004,
  title={Renormalization algorithms for Quantum-Many Body Systems in two and higher dimensions},
  author={Verstraete, Frank and Cirac, J. Ignacio},
  year={2004},
  eprint={cond-mat/0407066},
  archivePrefix={arXiv}
}

@article{Verstraete2006,
  title={Criticality, the Area Law, and the Computational Power of Projected Entangled Pair States},
  author={Verstraete, F. and Wolf, M. M. and Perez-Garcia, D. and Cirac, J. I.},
  journal={Physical Review Letters},
  volume={96},
  pages={220601},
  year={2006},
  doi={10.1103/PhysRevLett.96.220601}
}

@article{Cirac2021,
  title={Matrix product states and projected entangled pair states: Concepts, symmetries, theorems},
  author={Cirac, J. Ignacio and Perez-Garcia, David and Schuch, Norbert and Verstraete, Frank},
  journal={Reviews of Modern Physics},
  volume={93},
  pages={045003},
  year={2021},
  doi={10.1103/RevModPhys.93.045003}
}

@article{Eisert2010,
  title={Colloquium: Area laws for the entanglement entropy},
  author={Eisert, J. and Cramer, M. and Plenio, M. B.},
  journal={Reviews of Modern Physics},
  volume={82},
  pages={277--306},
  year={2010},
  doi={10.1103/RevModPhys.82.277}
}

@article{Vanderstraeten2022,
  title={Variational methods for contracting projected entangled-pair states},
  author={Vanderstraeten, Laurens and Burgelman, Lander and Ponsioen, Boris and Van Damme, Maarten and Vanhecke, Bram and Corboz, Philippe and Haegeman, Jutho and Verstraete, Frank},
  journal={Physical Review B},
  volume={105},
  pages={195140},
  year={2022},
  doi={10.1103/PhysRevB.105.195140}
}

@article{Fishman2018,
  title={Faster Methods for Contracting Infinite Two-Dimensional Tensor Networks},
  author={Fishman, Matthew T. and Vanderstraeten, Laurens and Zauner-Stauber, V. and Haegeman, Jutho and Verstraete, Frank},
  journal={Physical Review B},
  volume={98},
  pages={235148},
  year={2018},
  doi={10.1103/PhysRevB.98.235148}
}

@article{Schuch2007,
  title={Computational Complexity of Projected Entangled Pair States},
  author={Schuch, Norbert and Wolf, Michael M. and Verstraete, Frank and Cirac, J. Ignacio},
  journal={Physical Review Letters},
  volume={98},
  pages={140506},
  year={2007},
  doi={10.1103/PhysRevLett.98.140506}
}

@article{Haferkamp2020,
  title={Contracting projected entangled pair states is average-case hard},
  author={Haferkamp, Jonas and Hangleiter, Dominik and Eisert, Jens and Gluza, Marek},
  journal={Physical Review Research},
  volume={2},
  pages={013010},
  year={2020},
  doi={10.1103/PhysRevResearch.2.013010}
}

@article{Nishino1996,
  title={Corner Transfer Matrix Renormalization Group Method},
  author={Nishino, T. and Okunishi, K.},
  journal={Journal of the Physical Society of Japan},
  volume={65},
  pages={891--894},
  year={1996},
  doi={10.1143/JPSJ.65.891}
}

@article{Orus2009,
  title={Simulation of two-dimensional quantum systems on an infinite lattice revisited: Corner transfer matrix for tensor contraction},
  author={Or\'{u}s, Rom\'{a}n and Vidal, Guifr\'{e}},
  journal={Physical Review B},
  volume={80},
  pages={094403},
  year={2009},
  doi={10.1103/PhysRevB.80.094403}
}

@article{Lubasch2014,
  title={Algorithms for finite projected entangled pair states},
  author={Lubasch, Michael and Cirac, J. Ignacio and Ban\~{u}ls, Mari-Carmen},
  journal={Physical Review B},
  volume={90},
  pages={064425},
  year={2014},
  doi={10.1103/PhysRevB.90.064425}
}

@article{Orus2014,
  author  = {Or{\'u}s, Rom{\'a}n},
  title   = {A practical introduction to tensor networks: {Matrix} product states and projected entangled pair states},
  journal = {Annals of Physics},
  volume  = {349},
  pages   = {117--158},
  year    = {2014},
  doi     = {10.1016/j.aop.2014.06.013}
}

@article{PerezGarcia2007,
  title={Matrix product state representations},
  author={Perez-Garcia, David and Verstraete, Frank and Wolf, Michael M. and Cirac, J. Ignacio},
  journal={Quantum Information \& Computation},
  volume={7},
  number={5},
  pages={401--430},
  year={2007},
  eprint={quant-ph/0608197},
  archivePrefix={arXiv}
}

@article{White1992,
  title={Density matrix formulation for quantum renormalization groups},
  author={White, Steven R.},
  journal={Physical Review Letters},
  volume={69},
  pages={2863},
  year={1992},
  doi={10.1103/PhysRevLett.69.2863}
}

@article{Schollwock2011,
  title={The density-matrix renormalization group in the age of matrix product states},
  author={Schollw\"{o}ck, Ulrich},
  journal={Annals of Physics},
  volume={326},
  pages={96--192},
  year={2011},
  doi={10.1016/j.aop.2010.09.012}
}

@article{Yan2011,
  title={Spin-Liquid Ground State of the $S = 1/2$ Kagome Heisenberg Antiferromagnet},
  author={Yan, Simeng and Huse, David A. and White, Steven R.},
  journal={Science},
  volume={332},
  pages={1173--1176},
  year={2011},
  doi={10.1126/science.1201080}
}

@article{Bao2020,
  author  = {Bao, Yimu and Choi, Soonwon and Altman, Ehud},
  title   = {Theory of the phase transition in random unitary circuits with measurements},
  journal = {Physical Review B},
  volume  = {101},
  pages   = {104301},
  year    = {2020},
  doi     = {10.1103/PhysRevB.101.104301}
}

@article{Jian2020,
  author  = {Jian, Chao-Ming and You, Yi-Zhuang and Vasseur, Romain and Ludwig, Andreas W. W.},
  title   = {Measurement-induced criticality in random quantum circuits},
  journal = {Physical Review B},
  volume  = {101},
  pages   = {104302},
  year    = {2020},
  doi     = {10.1103/PhysRevB.101.104302}
}

@article{Dalibard1992,
  title={Wave-function approach to dissipative processes in quantum optics},
  author={Dalibard, Jean and Castin, Yvan and M{\o}lmer, Klaus},
  journal={Physical Review Letters},
  volume={68},
  pages={580--583},
  year={1992},
  doi={10.1103/PhysRevLett.68.580}
}

@article{Plenio1998,
  author  = {Plenio, Martin B. and Knight, Peter L.},
  title   = {The quantum-jump approach to dissipative dynamics in quantum optics},
  journal = {Reviews of Modern Physics},
  volume  = {70},
  pages   = {101--144},
  year    = {1998},
  doi     = {10.1103/RevModPhys.70.101}
}

@article{Molmer1993,
  author  = {M{\o}lmer, Klaus and Castin, Yvan and Dalibard, Jean},
  title   = {Monte {Carlo} wave-function method in quantum optics},
  journal = {Journal of the Optical Society of America B},
  volume  = {10},
  pages   = {524--538},
  year    = {1993},
  doi     = {10.1364/JOSAB.10.000524}
}

@book{Carmichael1993,
  author    = {Carmichael, Howard J.},
  title     = {An Open Systems Approach to Quantum Optics},
  publisher = {Springer},
  address   = {Berlin},
  series    = {Lecture Notes in Physics Monographs},
  volume    = {18},
  year      = {1993},
  doi       = {10.1007/978-3-540-47620-7}
}

@article{Kummerer2004,
  author  = {K{\"u}mmerer, Burkhard and Maassen, Hans},
  title   = {A pathwise ergodic theorem for quantum trajectories},
  journal = {Journal of Physics A: Mathematical and General},
  volume  = {37},
  number  = {49},
  pages   = {11889--11896},
  year    = {2004},
  doi     = {10.1088/0305-4470/37/49/11889}
}

@article{Cheng2023,
  author  = {Cheng, Zihan and Ippoliti, Matteo},
  title   = {Efficient Sampling of Noisy Shallow Circuits via Monitored Unraveling},
  journal = {PRX Quantum},
  volume  = {4},
  pages   = {040326},
  year    = {2023},
  doi     = {10.1103/PRXQuantum.4.040326}
}

@article{Browaeys2020,
  title={Many-body physics with individually controlled {Rydberg} atoms},
  author={Browaeys, Antoine and Lahaye, Thierry},
  journal={Nature Physics},
  volume={16},
  pages={132--142},
  year={2020},
  doi={10.1038/s41567-019-0733-z}
}

@article{Ebadi2021,
  title={Quantum phases of matter on a 256-atom programmable quantum simulator},
  author={Ebadi, Sepehr and others},
  journal={Nature},
  volume={595},
  pages={227--232},
  year={2021},
  doi={10.1038/s41586-021-03582-4}
}

@article{Graham2023MidCircuit,
  title={Mid-circuit measurements on a single-species neutral alkali atom quantum processor},
  author={Graham, T. M. and Phuttitarn, L. and Chinnarasu, R. and Song, Y. and Poole, C. and Jooya, K. and Scott, J. and Scott, A. and Eichler, P. and Saffman, M.},
  journal={arXiv preprint arXiv:2303.10051},
  year={2023},
  eprint={2303.10051},
  archivePrefix={arXiv}
}

@article{Bluvstein2024,
  title={Logical quantum processor based on reconfigurable atom arrays},
  author={Bluvstein, Dolev and others},
  journal={Nature},
  volume={626},
  pages={58--65},
  year={2024},
  doi={10.1038/s41586-023-06927-3}
}

@article{Blote2002,
  title={Cluster {Monte Carlo} simulation of the transverse {Ising} model},
  author={Bl{\"o}te, Henk W. J. and Deng, Youjin},
  journal={Physical Review E},
  volume={66},
  pages={066110},
  year={2002},
  doi={10.1103/PhysRevE.66.066110}
}

@article{HibatAllah2020,
  author={Hibat-Allah, Mohamed and Ganahl, Martin and Hayward, Lauren E. and Melko, Roger G. and Carrasquilla, Juan},
  title={Recurrent neural network wave functions},
  journal={Physical Review Research},
  volume={2},
  pages={023358},
  year={2020},
  doi={10.1103/PhysRevResearch.2.023358}
}

@article{Dagotto1989,
  author  = {Dagotto, Elbio and Moreo, Adriana},
  title   = {Phase diagram of the frustrated spin-$\frac{1}{2}$ {Heisenberg} antiferromagnet in 2 dimensions},
  journal = {Physical Review Letters},
  volume  = {63},
  pages   = {2148--2151},
  year    = {1989},
  doi     = {10.1103/PhysRevLett.63.2148}
}

@article{Haghshenas2018,
  author  = {Haghshenas, R. and Sheng, D. N.},
  title   = {{$U(1)$}-symmetric infinite projected entangled-pair state study of the spin-$\frac{1}{2}$ square {$J_1$--$J_2$} {Heisenberg} model},
  journal = {Physical Review B},
  volume  = {97},
  pages   = {174408},
  year    = {2018},
  doi     = {10.1103/PhysRevB.97.174408}
}

@article{Gong2014,
  title={Plaquette ordered phase and quantum phase diagram in the spin-$1/2$ $J_1$-$J_2$ square Heisenberg model},
  author={Gong, Shou-Shu and Zhu, Wei and Sheng, D. N. and Motrunich, Olexei I. and Fisher, Matthew P. A.},
  journal={Physical Review Letters},
  volume={113},
  pages={027201},
  year={2014},
  doi={10.1103/PhysRevLett.113.027201}
}

@article{Jiang2012,
  title={Spin liquid ground state of the spin-$1/2$ square $J_1$-$J_2$ Heisenberg model},
  author={Jiang, Hong-Chen and Yao, Hong and Balents, Leon},
  journal={Physical Review B},
  volume={86},
  pages={024424},
  year={2012},
  doi={10.1103/PhysRevB.86.024424}
}

@article{Nomura2021,
  title={Dirac-type nodal spin liquid revealed by refined quantum many-body solver using neural-network wave function, correlation ratio, and level spectroscopy},
  author={Nomura, Yusuke and Imada, Masatoshi},
  journal={Physical Review X},
  volume={11},
  pages={031034},
  year={2021},
  doi={10.1103/PhysRevX.11.031034}
}

@article{Liu2022,
  title={Gapless quantum spin liquid and global phase diagram of the spin-$1/2$ $J_1$-$J_2$ square antiferromagnetic Heisenberg model},
  author={Liu, Wen-Yuan and Gong, Shou-Shu and Li, Yu-Bin and Poilblanc, Didier and Chen, Wei-Qiang and Gu, Zheng-Cheng},
  journal={Science Bulletin},
  volume={67},
  pages={1077--1085},
  year={2022},
  doi={10.1016/j.scib.2022.03.010}
}

@article{Qian2024,
  title={Absence of spin liquid phase in the $J_1$-$J_2$ Heisenberg model on the square lattice},
  author={Qian, Xiangjian and Qin, Mingpu},
  journal={Physical Review B},
  volume={109},
  pages={L161103},
  year={2024},
  doi={10.1103/PhysRevB.109.L161103}
}

@incollection{Hansen2006,
  title={The CMA evolution strategy: A comparing review},
  author={Hansen, Nikolaus},
  booktitle={Towards a New Evolutionary Computation},
  pages={75--102},
  year={2006},
  publisher={Springer},
  doi={10.1007/3-540-32494-1_4}
}

@article{Kitaev2006,
  title={Topological Entanglement Entropy},
  author={Kitaev, Alexei and Preskill, John},
  journal={Physical Review Letters},
  volume={96},
  pages={110404},
  year={2006},
  doi={10.1103/PhysRevLett.96.110404}
}

@article{LevinWen2006,
  title={Detecting Topological Order in a Ground State Wave Function},
  author={Levin, Michael and Wen, Xiao-Gang},
  journal={Physical Review Letters},
  volume={96},
  pages={110405},
  year={2006},
  doi={10.1103/PhysRevLett.96.110405}
}

@article{Liu2019vqe,
  author  = {Liu, Jin-Guo and Zhang, Yi-Hong and Wan, Yuan and Wang, Lei},
  title   = {Variational Quantum Eigensolver with Fewer Qubits},
  journal = {Physical Review Research},
  volume  = {1},
  pages   = {023025},
  year    = {2019},
  doi     = {10.1103/PhysRevResearch.1.023025},
  eprint  = {1902.02663},
  archivePrefix = {arXiv}
}

@article{ITensor,
  title={The {ITensor} Software Library for Tensor Network Calculations},
  author={Fishman, Matthew and White, Steven R. and Stoudenmire, E. Miles},
  journal={SciPost Phys. Codebases},
  pages={4},
  year={2022},
  doi={10.21468/SciPostPhysCodeb.4},
  url={https://scipost.org/10.21468/SciPostPhysCodeb.4}
}

@misc{MPSKit,
  title        = {{MPSKit}.jl},
  author       = {Devos, Lukas and Van Damme, Maarten and Haegeman, Jutho},
  year         = {2025},
  howpublished = {\url{https://github.com/QuantumKitHub/MPSKit.jl}}
}

@misc{PEPSKit,
  title={PEPSKit.jl},
  author={Lootens, Lukas and others},
  year={2024},
  note={\url{https://github.com/QuantumKitHub/PEPSKit.jl}}
}

@techreport{Rong2026IsoPEPSData,
  author      = {Rong, Yuqing},
  title       = {Data for: Sampling Isometric Tensor Network States with Monitored Quantum Circuits},
  type        = {Dataset},
  number      = {Version 1.0.0},
  institution = {Zenodo},
  year        = {2026},
  doi         = {10.5281/zenodo.22163225},
  url         = {https://doi.org/10.5281/zenodo.22163225}
}

\appendix 
\section{Equivalence between cylindrical and spiral PEPS}
\label{app:spiral_equivalence}

Figure~\ref{fig:spiral_equivalence} illustrates the index deformation established below.

\begin{samepage}
\begin{inlinefigurecontent}
  \includegraphics[width=0.88\textwidth]{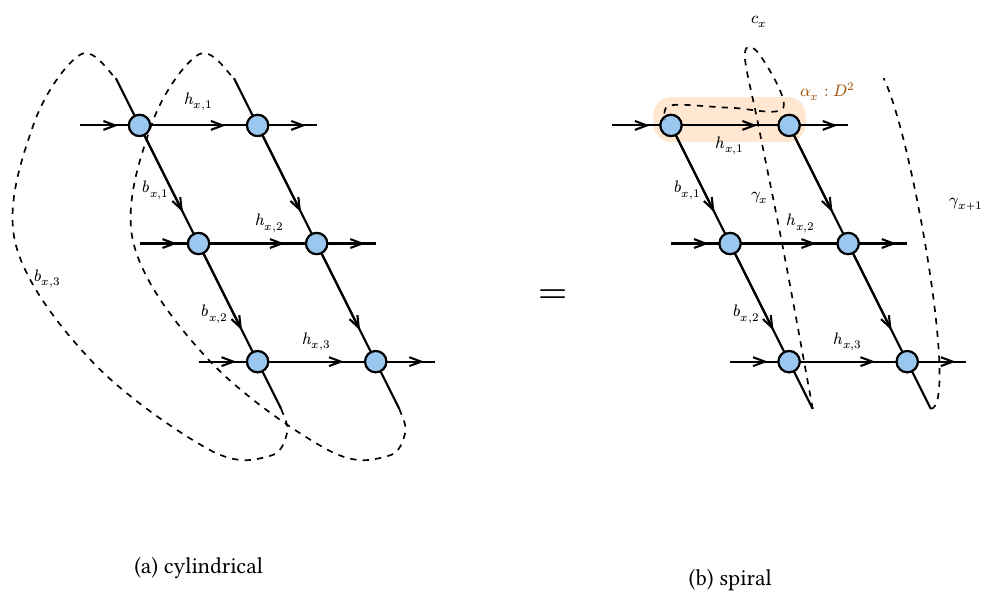}
  \caption{Equivalence between (a) cylindrical and (b) spiral PEPS for $W=3$.
  The closing bond $b_{x,3}$ is deformed into the spiral wire $\gamma_x$ folded
  back as $c_x$ (the bend being the Kronecker $\delta$); $c_x$ fuses with the
  first-row bond $h_{x,1}$ into the enlarged bond $\alpha_x$ of dimension $D^2$.}
  \label{fig:spiral_equivalence}
\end{inlinefigurecontent}
\end{samepage}

For clarity, consider circumference $W=3$, with sites labeled by $(x,y)$,
$y=1,2,3$.  In the ordinary cylindrical PEPS, let $h_{x,y}$ connect
$(x,y)$ to $(x+1,y)$ and let $b_{x,1}$, $b_{x,2}$, and $b_{x,3}$ connect
$(x,1)$ to $(x,2)$, $(x,2)$ to $(x,3)$, and $(x,3)$ back to $(x,1)$,
respectively.  The amplitude of a spin configuration is
\begin{equation}
\begin{aligned}
\psi_{\rm cyl}(\{s\})
=
\sum_{\{h,b\}}\prod_x
&
A_{x,1}^{s_{x,1}}(h_{x-1,1},h_{x,1};b_{x,3},b_{x,1}) \\
&\times
A_{x,2}^{s_{x,2}}(h_{x-1,2},h_{x,2};b_{x,1},b_{x,2}) \\
&\times
A_{x,3}^{s_{x,3}}(h_{x-1,3},h_{x,3};b_{x,2},b_{x,3}) .
\end{aligned}
\label{eq:cyl_peps_coeff}
\end{equation}

The spiral ordering replaces the closing link within column $x$ by a link
\[
\gamma_x:(x,3)\to(x+1,1).
\]
To keep the contraction unchanged, we enlarge the first-row horizontal index,
\begin{equation}
\alpha_x=(h_{x,1},c_x),\qquad \dim \alpha_x=D^2,
\label{eq:alpha_enlargement}
\end{equation}
where $c_x$ carries the original closing-bond index.  All other virtual indices
retain dimension $D$.  Writing
$\alpha_{x-1}=(h_{x-1,1},c_{x-1})$ and
$\alpha_x=(h_{x,1},c_x)$, define the spiral tensors as follows, with the
left-hand sides carrying virtual arguments
$(\alpha_{x-1},\alpha_x;\gamma_{x-1},b_{x,1})$,
$(h_{x-1,2},h_{x,2};b_{x,1},b_{x,2})$, and
$(h_{x-1,3},h_{x,3};b_{x,2},\gamma_x)$, respectively:
\begin{equation}
\begin{aligned}
B_{x,1}^{s_{x,1}}
&=
\delta_{\gamma_{x-1},c_{x-1}}\,
A_{x,1}^{s_{x,1}}(h_{x-1,1},h_{x,1};c_x,b_{x,1}),\\
B_{x,2}^{s_{x,2}}
&=
A_{x,2}^{s_{x,2}}(h_{x-1,2},h_{x,2};b_{x,1},b_{x,2}),\\
B_{x,3}^{s_{x,3}}
&=
A_{x,3}^{s_{x,3}}(h_{x-1,3},h_{x,3};b_{x,2},\gamma_x).
\end{aligned}
\label{eq:spiral_tensors}
\end{equation}
Thus $B_{x,1}$ uses $c_x$ as the closing index entering the top tensor in
column $x$, while its Kronecker delta enforces that the incoming spiral bond
$\gamma_{x-1}$ equals the closing index $c_{x-1}$ stored in the previous
enlarged horizontal bond.

The spiral coefficient is obtained by summing the product of the three
$B$ tensors in each column over $\alpha$, $h$, $b$, and $\gamma$.
Because summing over $\alpha_x$ is the same as summing over $h_{x,1}$ and
$c_x$, and the delta functions set $\gamma_x=c_x$ for every column, performing the
$\gamma$ sums gives
\begin{equation}
\begin{aligned}
\psi_{\rm sp}(\{s\})
=
\sum_{\{h,b,c\}}\prod_x
&
A_{x,1}^{s_{x,1}}(h_{x-1,1},h_{x,1};c_x,b_{x,1})\\
&\times
A_{x,2}^{s_{x,2}}(h_{x-1,2},h_{x,2};b_{x,1},b_{x,2})\\
&\times
A_{x,3}^{s_{x,3}}(h_{x-1,3},h_{x,3};b_{x,2},c_x).
\end{aligned}
\end{equation}
Relabeling $c_x$ as $b_{x,3}$ gives exactly
Eq.~\eqref{eq:cyl_peps_coeff}.  Hence
\begin{equation}
|\psi_{\rm sp}\rangle=|\psi_{\rm cyl}\rangle .
\end{equation}
The spiral geometry is therefore an index reshuffling of the cylindrical PEPS:
only one horizontal bond, here the first-row bond, must be enlarged from $D$ to
$D^2$ to carry the original closing index.
The same construction extends directly to circumference $W$ by carrying the
closing bond $b_{x,W}$ in the enlarged first-row horizontal index and replacing
it with the spiral link $\gamma_x:(x,W)\to(x+1,1)$.

\section{TFIM diagnostics}
\label{app:tfim}

This appendix collects diagnostics that support the TFIM benchmark in Sec.~\ref{sec:results}. We first characterize thermalization and finite-shot convergence, then examine the correlation length, local correlations, and magnetization across the field range. We finally test the sensitivity of the sampled energy estimator to a simple readout-noise model.

\subsection{Energy dynamics and sampling convergence}
Figure~\ref{fig:tfim_energy} shows the thermalization of the ensemble-averaged local energy from a common initial boundary state. The approach to a stable plateau provides an operational diagnostic for choosing the thermalization length $l_{\mathrm{th}}$: the initial columns are discarded until the curves no longer display a systematic transient.
Figure~\ref{fig:tfim_variance_convergence} presents a representative finite-shot convergence analysis for the TFIM at $g=3.0$, showing the expected $1/N$ decrease of the estimator variance.

\begin{inlinefigurecontent}
  \subfigure[]{%
    \includegraphics[width=0.48\columnwidth]{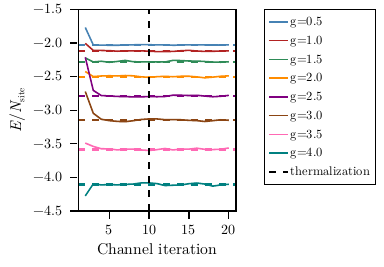}%
    \label{fig:tfim_energy}%
  }
  \hfill
  \subfigure[]{%
    \includegraphics[width=0.48\columnwidth]{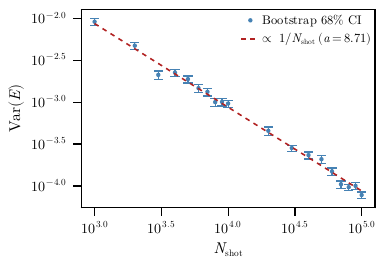}%
    \label{fig:tfim_variance_convergence}%
  }
  \caption{%
  \textbf{Energy dynamics and sampling convergence for the TFIM.}
  (a)~Ensemble-averaged instantaneous local energy versus channel iteration $l$ for representative transverse fields $g$, obtained from $M=10000$ independent monitored trajectories with circumference $W=3$ and bond dimension $D=2$. The dashed horizontal lines show the energy density obtained by exact tensor-network contraction of the corresponding optimized states, and the vertical dashed line marks the plateau-onset diagnostic for choosing a thermalization cutoff.
  (b)~Variance of the energy estimator at $g=3.0$ versus sample count $N$, obtained from $200$ bootstrap resamples. The dashed line is the fit $\operatorname{Var}(E)=a/N$ with $a=8.71$.
  }
  \label{fig:tfim_energy_sampling}
\end{inlinefigurecontent}

\subsection{Correlation and magnetization diagnostics}
Figure~\ref{fig:tfim_corr_length_scaling} shows that the transfer-matrix correlation length grows as the field approaches the crossover region and decreases at larger $g$. The maximum is shifted and broadened relative to the bulk critical point $g_c\simeq3.04438(2)$. This difference is expected for the finite circumference $W=3$ and modest bond dimension $D=2$, so the curve is used as a qualitative correlation diagnostic rather than a precise estimate of the thermodynamic critical point.
Figure~\ref{fig:tfim_connected_corr} complements the transfer-matrix analysis with directly sampled, short-range observables. Together with Fig.~\ref{fig:tfim_corr_length_scaling}, it shows that the crossover is visible in both the transfer spectrum and local real-space correlations.
Figure~\ref{fig:tfim_magnetization} shows a smooth crossover from longitudinal polarization, measured by $|Z|$, to transverse polarization, measured by $|X|$, as the field increases.

\newpage
\begin{samepage}
\begin{inlinefigurecontent}
  \subfigure[]{%
    \includegraphics[width=0.32\columnwidth]{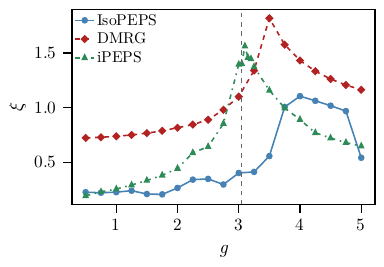}%
    \label{fig:tfim_corr}%
    \label{fig:tfim_corr_length_scaling}%
  }
  \hfill
  \subfigure[]{%
    \includegraphics[width=0.32\columnwidth]{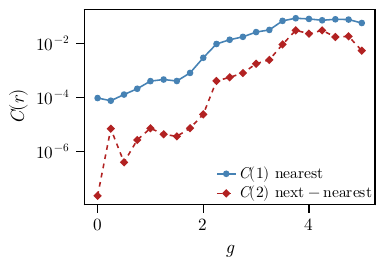}%
    \label{fig:tfim_connected_corr}%
  }
  \hfill
  \subfigure[]{%
    \includegraphics[width=0.32\columnwidth]{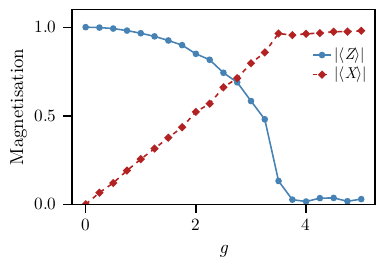}%
    \label{fig:tfim_magnetization}%
  }
  \caption{%
  \textbf{Correlation and magnetization diagnostics across the TFIM crossover.}
  (a)~Correlation length $\xi$ for circumference $W=3$, extracted from the transfer-matrix spectrum of the sample-optimized isoPEPS and compared with DMRG on the same cylinder and an infinite-PEPS reference obtained using PEPSKit.jl~\cite{PEPSKit}. The vertical dotted line marks the bulk square-lattice critical point $g_c \simeq 3.04438(2)$~\cite{Blote2002}.
  (b)~Connected correlations $C(r)=\langle Z_i Z_j \rangle-\langle Z_i\rangle\langle Z_j\rangle$ for nearest-neighbor pairs, $C(1)$, and next-nearest-neighbor pairs, $C(2)$.
  (c)~Longitudinal and transverse magnetizations: $|Z|$ decreases as $g$ increases, while $|X|$ grows toward saturation.
  }
  \label{fig:tfim_correlation_magnetization}
\end{inlinefigurecontent}
\end{samepage}

\subsection{Readout-noise sensitivity}
\label{app:tfim_readout}
To estimate sensitivity to measurement errors, we independently flip each recorded measurement bit with probability $p$ before constructing the energy estimator. At $g=3.0$, Fig.~\ref{fig:tfim_readout_energy} shows that this symmetric readout-error model biases the measured energy density upward. Figure~\ref{fig:tfim_readout_energy_error} shows that the corresponding absolute deviation grows approximately linearly over the tested range $p=0\%$--$5\%$. This calculation characterizes noise sensitivity only; no readout-error mitigation is applied.

\begin{inlinefigurecontent}
  \subfigure[]{%
    \includegraphics[width=0.48\columnwidth]{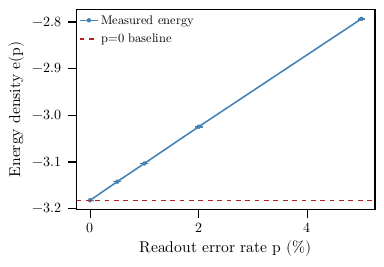}%
    \label{fig:tfim_readout_energy}%
  }
  \hfill
  \subfigure[]{%
    \includegraphics[width=0.48\columnwidth]{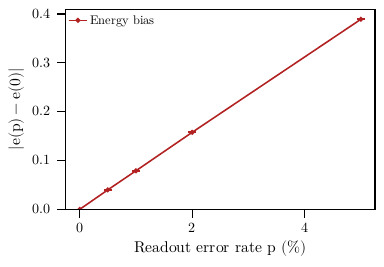}%
    \label{fig:tfim_readout_energy_error}%
  }
  \caption{%
  \textbf{Readout-noise sensitivity of the TFIM energy estimator at $g=3.0$.}
  (a)~Measured energy density $e(p)$ when each recorded measurement bit is independently flipped with probability $p$. The dashed red line marks the zero-error baseline $e(0)$; increasing readout error shifts the estimated energy to higher values.
  (b)~Absolute deviation $|e(p)-e(0)|$ from the zero-error energy density. The deviation grows approximately linearly over the tested range $p=0\%$--$5\%$.
  }
  \label{fig:tfim_readout}
\end{inlinefigurecontent}

\section[Heisenberg J1--J2 energy]{Heisenberg $J_1$--$J_2$}
\label{app:j1j2}

\subsection{Sampling variance}
For the Heisenberg model at $J_2/J_1=0.5$, we quantify the sampling uncertainty of the energy estimator using $200$ bootstrap resamples for each shot count $N_{\mathrm{shot}}$. Figure~\ref{fig:j1j2_variance} shows the expected approximate $1/N_{\mathrm{shot}}$ decrease of the variance. The fitted prefactor sets the sample count required to reach a target statistical precision.

\begin{inlinefigurecontent}
  \includegraphics[width=0.7\columnwidth]{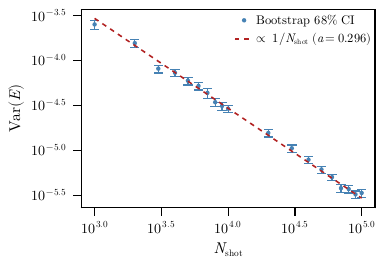}
  \caption{%
  \textbf{Sampling variance for the $J_1$--$J_2$ Heisenberg model at $J_2/J_1=0.5$.}
  Variance of the energy estimator versus shot count $N_{\mathrm{shot}}$, obtained from $200$ bootstrap resamples. The dashed line is the fit $\operatorname{Var}(E)=a/N$ with $a=0.296$.
  }
  \label{fig:j1j2_variance}
\end{inlinefigurecontent}

\subsection{Energy-density benchmark}
Figure~\ref{fig:heisenberg_energy_vs_j2} compares the $D=2$ isoPEPS energy density with DMRG results at bond dimensions $D=2$ and $D=192$. The isoPEPS curve has the same broad trend as the $D=2$ DMRG result across the scanned frustration range but remains above the more converged $D=192$ energy, with the largest difference in the intermediate frustrated regime. The comparison therefore indicates that finite bond dimension is a significant limitation of the present ansatz and should be accounted for when interpreting its quantitative accuracy.

\begin{inlinefigurecontent}
  \includegraphics[width=0.7\columnwidth]{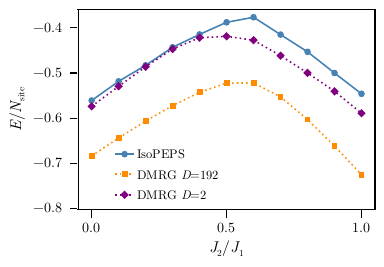}
  \caption{\textbf{Energy per site versus frustration ratio $J_2/J_1$ for the Heisenberg model.}
  Energy densities of the $D=2$ isoPEPS ansatz and DMRG calculations with bond dimensions $D=2$ and $D=192$. The isoPEPS result shows the same broad trend as the $D=2$ reference, while the lower $D=192$ energy displays the remaining finite-bond-dimension error, most clearly in the frustrated intermediate regime.}
  \label{fig:heisenberg_energy_vs_j2}
\end{inlinefigurecontent}

\subsection{Real-space bond correlations}
Figure~\ref{fig:j1j2_bond_correlations} shows the nearest-neighbor spin correlations obtained by the optimized $D=2$ isoPEPS. At $J_2/J_1=0$, antiferromagnetic correlations occur on both bond orientations, consistent with N\'eel order. In the intermediate regime $J_2/J_1=0.5$--$0.6$, the strongest antiferromagnetic correlations form alternating vertical bonds, revealing the columnar dimer pattern probed by $M_D^2(0,\pi)$. At $J_2/J_1=1$, vertical bonds remain antiferromagnetic while horizontal bonds become ferromagnetic, consistent with stripe correlations.

\begin{inlinefigurecontent}
  \includegraphics[width=\columnwidth]{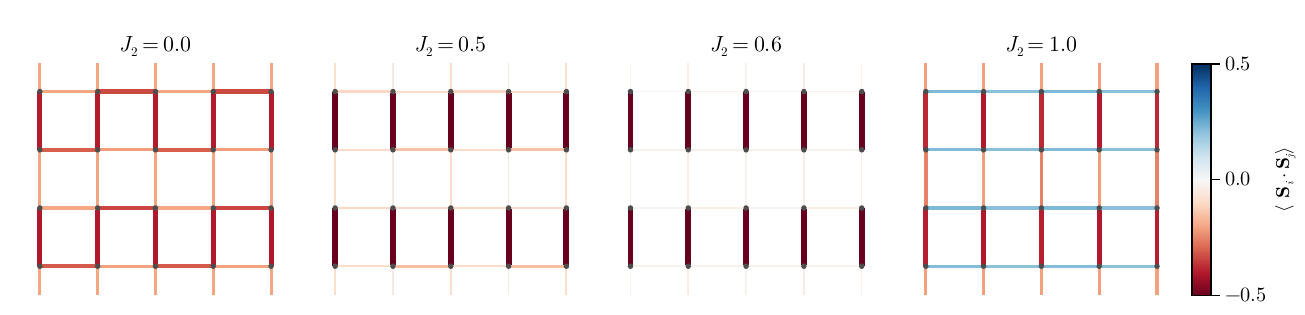}
  \caption{\textbf{Real-space nearest-neighbor spin correlations in the Heisenberg $J_1$--$J_2$ model.}
  Exact-contraction values of $\langle \mathbf S_i\!\cdot\!\mathbf S_j\rangle$ for the optimized $D=2$ isoPEPS at $J_2/J_1=0.0$, $0.5$, $0.6$, and $1.0$, with $J_1=1$. The bond colors encode the sign and magnitude of the correlations. Alternating strong antiferromagnetic vertical bonds appear at $J_2/J_1=0.5$--$0.6$, while the small- and large-frustration panels display N\'eel- and stripe-like patterns, respectively.}
  \label{fig:j1j2_bond_correlations}
\end{inlinefigurecontent}

\end{document}